\documentclass{emulateapj}

\usepackage{graphicx}
\usepackage{amsmath}
\usepackage{natbib}
\usepackage{amssymb}

\usepackage[breaklinks,colorlinks,urlcolor=blue,citecolor=blue,linkcolor=blue]{hyperref}
\usepackage{hyperref}

\newcommand{\be}{\begin{eqnarray}}
\newcommand{\ee}{\end{eqnarray}}

\newcommand{\lp}{\left(}
\newcommand{\rp}{\right)}
\newcommand{\lb}{\left[}
\newcommand{\rb}{\right]}

\newcommand{\moon}{{\rm M}}

\newcommand{\slugcom}{Accepted for publication in The Astronomical Journal}
\slugcomment{\slugcom}

\begin{document}

\normalsize

% -----------------------------------------------------------
% -----------------------------------------------------------

\title{Exoplanets Torqued by the Combined Tides of a Moon and Parent Star}

\author{Anthony L. Piro}

\affil{The Observatories of the Carnegie Institution for Science, 813 Santa Barbara St., Pasadena, CA 91101, USA; piro@carnegiescience.edu}

\begin{abstract}

In recent years, there has been interest in Earth-like exoplanets in the habitable zones of low mass stars \mbox{($\sim0.1-0.6\,M_\odot$)}. Furthermore, it has been argued that a large moon may be important for stabilizing conditions on a planet for life. If these two features are combined, then an exoplanet can feel a similar tidal influence from both its moon and parent star, leading to potentially interesting dynamics. The moon's orbital evolution depends on the exoplanet's initial spin period $P_0$.  When $P_0$ is small, transfer of the exoplanet's angular momentum to the moon's orbit can cause the moon to migrate outward sufficiently to be stripped by the star. When $P_0$ is large, the moon migrates less and the star's tidal torques spin down the exoplanet. Tidal interactions then cause the moon to migrate inward until it is likely tidally disrupted by the exoplanet and potentially produces rings. While one may think that these findings preclude the presence of moons for the exoplanets of low mass stars, in fact a wide range of timescales are found for the loss or destruction of the moon; it can take $\sim10^6-10^{10}\,{\rm yrs}$ depending on the system parameters. When the moon is still present, the combined tidal torques force the exoplanet to spin asynchronously with respect to both its moon and parent star, which tidally heats the exoplanet. This can produce heat fluxes comparable to those currently coming through the Earth, arguing that combined tides may be a method for driving tectonic activity in exoplanets.

\end{abstract}

	
\section{Introduction}
\label{sec:introduction}

It has been revealed in recent years that terrestrial extrasolar planets are basically ubiquitous in our Galaxy \citep{Burke15,Mulders15}, and, although extrasolar moons have yet to be discovered, certainly many of these exoplanets have moons just like the planets in our own solar system. Whether or not a planet has a moon is not just a minor curiosity, but potentially fundamental to the question of whether these planets host life. In the case of the Earth, the Moon's relatively large size allows it to stabilize the Earth's obliquity. This may be crucial for stabilizing conditions on Earth for a sufficiently long time to allow life to develop (\citealp{Ward00}, although also see \citealp{Lissauer12,Li14}).

Another factor that may affect a planet's ability to host life is the presence of internal heating and volcanism. The associated tectonic activity due to the movement of plates sitting atop a fluid mantle can trap atmospheric gases such as carbon dioxide into rocks and help stabilize the climate \citep{Walker81,Sleep01,Foley16}. Tectonic activity can also cycle fresh rock and minerals out of deeper regions of the planet, providing the building blocks and nutrients for life. Heating can provide the energy needed to drive biochemical reactions as is seen from hydrothermal vents on Earth. In the case of the Earth, this heating is driven by a combination of radioactivity, latent heat, and heat from formation, but in principle a moon could also provide a source of heating through tidal interactions, as is seen for Io. The easiest way to drive such tidal heating would be for a moon via an eccentric orbit. In the absence of a large eccentricity though there could also be cases where a planet is being tidally torqued by both its moon and parent star. The competing torques ensure that the planet is never perfectly synchronized to either the moon or star, so that tidal heating can continue to persist. 

Motivated by these possibilities, here I consider the evolution of a planet tidally torqued to a similar level by both its moon and parent star. Particular focus is on stars in the mass range of $M_*\approx 0.1-0.6\,M_\odot$, since they have two attractive properties: (1) there has been strong interest in studying exoplanets in the habitable zones of these stars, and (2) their habitable zones are closer in where tidal interactions with the star are more important. Of course various aspects of star-planet-moon tidal interactions have been investigated previously. As early as in \citet{Counselman73}, it was noted such systems would evolve toward one of three states, (1) the moon migrates inward until it reaches the planet, (2) the moon migrates outward until it escapes from the planet, and (3) the moon finds a stable state where its orbital frequency and the planetary spin frequency are at a mutual resonance. {Since this work, the problem has continued to be studied in various ways over a wide variety of possible systems \citep[e.g.,][]{Ward73,Touma94,Neron97,Barnes02,Sasaki12,Sasaki14,Adams16}.} Here I focus on particular aspects of this previous work by highlighting the possibility of tidal heating and the importance of the initial spin period of the planet in determining the resulting dynamics.

In Section \ref{sec:stage}, I motivate and summarize the parameter range for the star-planet-moon systems I will be considering. In Section \ref{sec:equations}, I present the set of equations I use to solve for the orbits and tides. In Section \ref{sec:timescales}, I summarize the characteristic timescale of the problem and solve for the evolution of the exoplanet's spin in the limit of no orbital evolution. This provides some intuition that is useful for solving the more detailed evolution of the system. In Section \ref{sec:solutions}, I explore the full evolution of the star-planet-moon system over a range of parameters, and I conclude in Section \ref{sec:conclusions} with a summary.

\section{Setting the Stage}
\label{sec:stage}

\begin{figure}
\begin{center}
  \includegraphics[width=0.47\textwidth,trim=0.0cm 0.0cm 0.0cm 0.0cm]{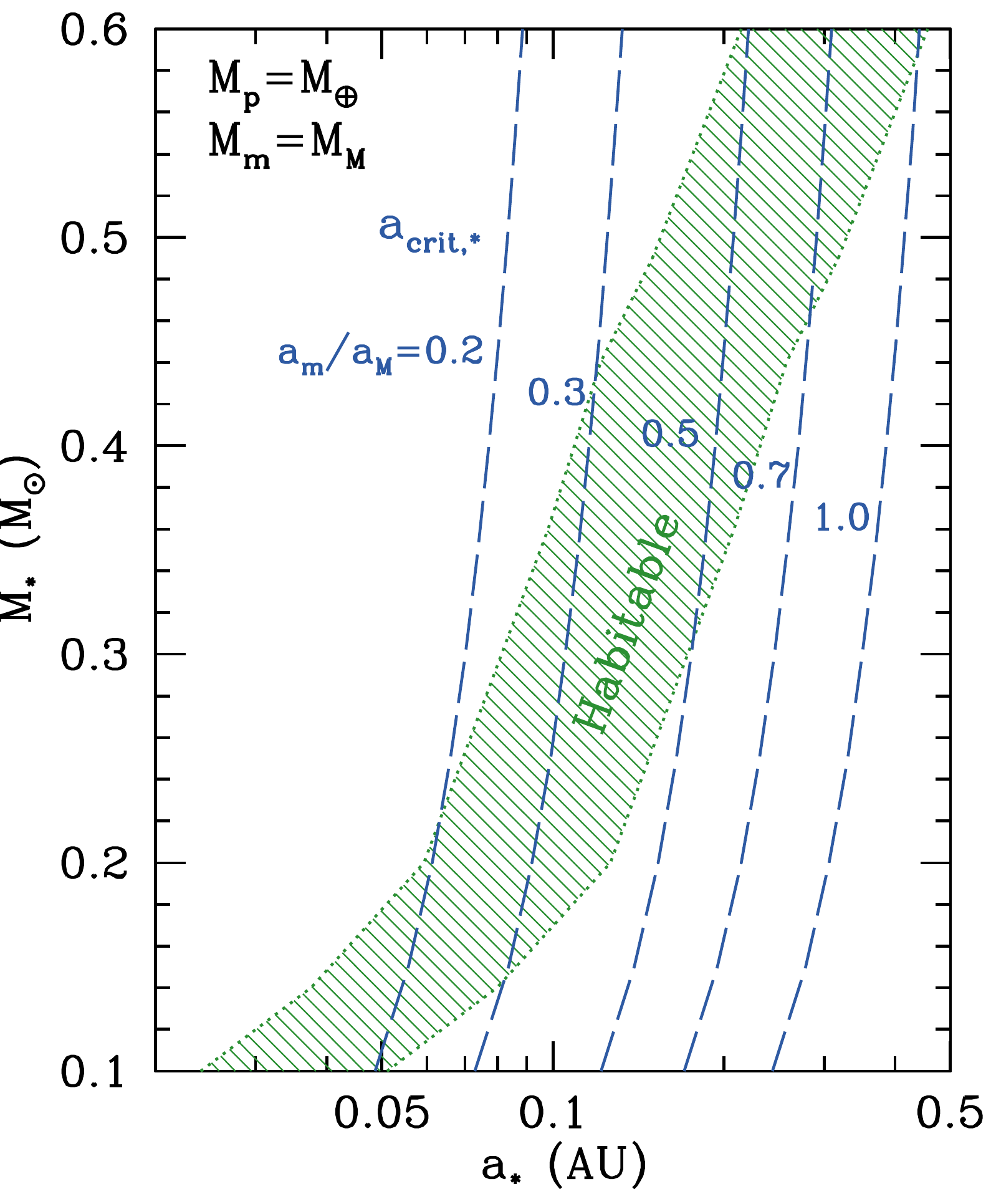}
  \end{center}
\caption{The habitable zone for stars in the mass range of $M_*=0.1-0.6\,M_\odot$ (green shaded region). This is defined as the range of distances which match the same flux received at distances of $0.8-1.7\,{\rm AU}$ from our Sun. Blue dashed lines indicate $a_{{\rm crit},*}$, the critical distance for the planet from the star, within which the moon would be stripped away. These are shown for different values of $a_m/a_\moon$ as labeled. In all cases I use \mbox{$M_p=M_\oplus$} and $M_m=M_\moon$.}
\epsscale{1.0}
\label{fig:timescales}
 \end{figure}

Before diving into the details of tidal interactions, it is helpful to motivate the range of parameters that will be considered. In Figure  \ref{fig:timescales}, I plot the habitable zone for stars in the mass range of $M_*=0.1-0.6\,M_\odot$ (green shaded region). This is estimated as the range of distances at which an exoplanet would receive the same flux as at distances of  $0.8-1.7\,{\rm AU}$ from the Sun {(this is the more optimistic range from the work of \citealp{Kasting93}--a less optimistic range would be $0.95-1.4\,{\rm AU}$).} In comparison to the Sun, the habitable zone must be fairly close to the parent star because the luminosity varies strongly with mass.

If an exoplanet with an exomoon is within this range of distances from the star, it is possible that the star's gravity is sufficiently strong to unbind that moon. The critical distance within which a moon can remain is proportional to the so-called Hill sphere, $R_{\rm H}$, so that
\be
	a_{{\rm crit},m} = fR_{\rm H} = f a_*\lp\frac{M_p}{3M_*} \rp^{1/3},
	\label{eq:roche}
\ee
for which I use a constant factor $f=0.49$ {\citep[][as appropriate for prograde orbits]{Domingos06} for this work \citep[although note that a value of $f=0.36$ has also been argued for in the past by][]{Holman99}.}  This relation can be inverted to find a critical distance of the planet from the star, within which the moon would be stripped away. This is given by
\be
	a_{{\rm crit},*}& =& \frac{a_m}{f} \lp\frac{3M_*}{M_p} \rp^{1/3}
	\nonumber
	\\
	&=& 0.52 
		\lp \frac{a_m}{a_\moon} \rp
		\lp \frac{M_p}{M_\oplus} \rp^{-1/3}
	\lp \frac{M_*}{M_\odot} \rp^{1/3}
	{\rm AU},
\ee
where $a_\moon=3.84\times10^{10}\,{\rm cm}$ is the distance between the Earth and Moon. In Figure  \ref{fig:timescales}, I plot $a_{{\rm crit},*}$ for different values of $a_m/a_\moon$ (blue dashed lines). In all cases, I assume the mass of the exoplanet is $M_p=M_\oplus$, where $M_\oplus=5.97\times10^{27}\,{\rm g}$ is the mass of the Earth, and the mass of the exomoon is $M_m=M_\moon$, where $M_\moon=7.35\times10^{25}\,{\rm g}$ is the mass of the Moon. From comparing these critical distances to the habitable zone, one can see that the moon must be sufficiently close to the exoplanet to prevent being stripped. For example, for $M_*=0.5\,M_\odot$ and $a_*=0.3\,{\rm AU}$, then the moon must be at a distance $a_m\lesssim0.7\,a_\moon$ from the exoplanet to stay bound. These considerations roughly set the parameters I will be using for the initial conditions of the dynamical evolution calculations.

\section{Basic Equations}
\label{sec:equations}

I next present the set of equations I will be using to solve for the dynamics of the planet's spin and the orbital separations. The basic strategy is to focus on the secular evolution of the star-planet-moon system rather than follow each orbit individually.  In this sense, the conservations equations used represent averages over many orbits. This allows the evolution of these systems to be considered over much longer timescales of $\sim10^5-10^{10}\,{\rm yrs}$ rather than being focused on the much shorter timescale of the orbits themselves. Further simplifications include assuming that the spin angular momentum of the planet is parallel to the orbital angular momentum of both the moon and planet.

\begin{figure*}
\begin{center}
  \includegraphics[width=0.8\textwidth,trim=0cm 3.0cm 1.0cm 2.0cm]{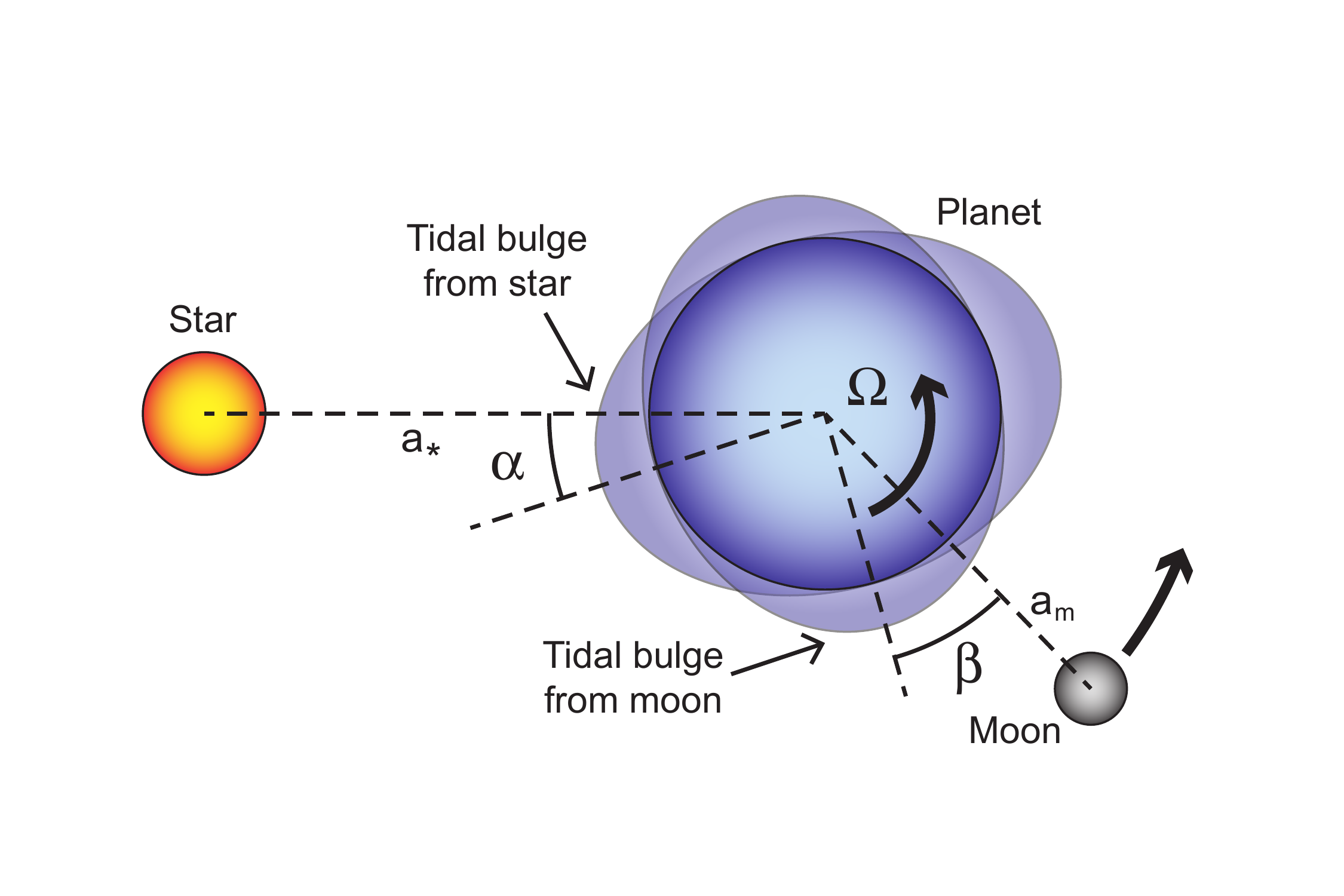}
  \end{center}
  \caption{Schematic showing the combined tidal effects on an planet from both a star and moon (definitely not drawn to scale). Two tidal bulges are raised by each the star and moon, which  proceed or lag behind the actual position of the forcing body depending on the time for the planet to react to the tides (parameterized by the time lag factor $\tau$) and the relative frequencies of the planet's spin and the orbits. In this specific example $n_m>\Omega>n_*$, so that the bulge from the star proceeds ahead of the star's position by an angle $\alpha$, while the bulge from the moon lags behind the moon's position by an angle $\beta$.}
  \label{fig:diagram}
\end{figure*}

Consider the configuration of a star, planet, and moon, with masses $M_*$, $M_p$, and $M_m$, respectively, as shown schematically in Figure \ref{fig:diagram}. The semi-major axis of the orbit of the star and planet is $a_*$ and the semi-major axis of the planet and moon is $a_m$. The orbits are all assumed to be circular. Tidal forces from the star and moon each generate two bulges on the planet on opposite sides. The frequency of these bulges in the frame of the planet are $\hat{\omega}_* = 2(n_*-\Omega)$ and $\hat{\omega}_m = 2(n_m-\Omega)$ due to the star and moon, respectively, where the factor of $2$ represents the two bulges, $\Omega$ is the spin of the planet, and the orbital frequencies are $n_*^2 = GM_*/a_*^3$ and  $n_m^2 = GM_p/a_m^3$.

These bulges are not perfectly aligned with the position of the mass generating the bulge because the planet is not a perfect fluid. Instead, the bulge can proceed or lag behind. This is emphasized in Figure \ref{fig:diagram}, where the bulge caused by the star proceeds the star by an angle $\alpha$ and the bulge caused by the moon lags behind by an angle $\beta$ (note that these angle are exaggerated here and are much smaller in real systems). These angles depend on the ``time lag'' of the tidal forcing of the planet $\tau$ and tidal forcing frequencies such that $\alpha \propto \tau \hat{\omega}_*$ and $\beta \propto \tau \hat{\omega}_m$. In Figure \ref{fig:diagram}, it is assumed that the planet is spinning faster than its orbit around the star, thus $\hat{\omega}_*<0$ and the bulge from the star proceeds its position.

Due to the misalignment between the bulge and the object causing the tide, the planet's spin can be torqued up or down. These torques over secular timescales can be approximated as  \citep{Ogilvie14}
\be
	N_* =  \frac{3}{2}k_2\tau \hat{\omega}_*\frac{GM_*^2R_p^5}{a_*^6},
\ee
and
\be
	N_m =\frac{3}{2}k_2\tau\hat{\omega}_m\frac{GM_m^2R_p^5}{a_m^6},
\ee
 from the star and moon, respectively, where $R_p$ is the radius of the planet and $k_2$ is the Love number which represents the planet's rigidity. In the specific case of Figure \ref{fig:diagram}, $N_*<0$ and $N_m>0$ because of whether the bulge proceeds of lags. These torques can be rewritten in terms of the orbital frequencies as
\be
	N_* = \frac{3}{2}k_2\tau \hat{\omega}_*M_*R_p^2 \lp\frac{R_p}{a_*} \rp^3 n_*^2,
\ee
and
\be
	N_m = \frac{3}{2}k_2\tau\hat{\omega}_m M_m R_p^2  \lp\frac{M_m}{M_p} \rp \lp\frac{R_p}{a_m} \rp^3 n_m^2.
\ee
Note that because I am assuming that $M_p\gg M_m$, there is an additional factor of $M_m/M_p$ for $N_m$.

Note that for this work I express the tidal torque using a {\em constant time lag model} \citep{Mignard79,Mignard80,Mignard81,Hut81,Heller11} rather than a {\em constant geometric lag model} \citep{MacDonald64,Goldreich66b,Murray99}, which would be parameterized with a quality factor $Q$. The relation between the two is $k_2\tau\hat{\omega} = \sigma k_2/Q$ with $\sigma = {\rm sgn}\,\hat{\omega}$ (see \citealp{Efroimsky13} for a discussion of the limitations and implications of different tidal implementations). I choose this formalism simply because it provides a more smooth transition from a positive torque (when the planet is spinning slowly) to zero torque (when the planet is tidal locked) to a negative torque (when the planet is spinning quickly). In contrast, if I leave $Q$ fixed, the torque changes more abruptly as the planet's spin evolves, which causes numerical issues when evolving the orbits and spins.

Another issue is that I consider the torque by the star from the star's bulge and the torque by the moon from the moon's bulge but not vice versa (for example, the torque by the star from the moon's bulge). This is a reasonable approximation for the secular limit, because over long timescales the moon's bulge, for example, will appear at all different positions with respect to the star and therefore will average to zero torque. If the orbits were followed individually, there would be more complicated behavior on shorter timescales, but this is beyond the secular focus of the present study.

The torques change the spin angular momentum of the planet as
\be
	\frac{d}{dt}(I_p \Omega) = N_*+ N_m,
	\label{eq:angular momentum}
\ee
where $I_p$ and $\Omega$ are the moment of inertia and spin frequency of the planet, respectively. To conserve angular momentum, the orbital separations must also change. The differential equations that describe these are
\be
	\frac{d}{dt}\lb M_p(GM_* a_*)^{1/2} \rb = - N_*,
	\label{eq:planet orbit}
\ee
and
\be
	\frac{d}{dt}\lb M_m(GM_p a_m)^{1/2} + I_mn_m \rb = - N_m,
	\label{eq:moon orbit}
\ee
where $I_m$ is the moment of inertia for the moon.  For simplicity, I assume that the moon is tidally locked to the planet, which gives rise to the term $I_mn_m$ in \mbox{Equation (\ref{eq:moon orbit}).} Since relatively little angular momentum is in the moon's spin, whether or not this tidal locking occurs does not impact my main conclusions. For the star, I assume that its spin is not changing appreciably from the small tidal torque of the planet and that it can be ignored.

From the above differential equations a number of key timescales can be identified. Rewriting Equation~(\ref{eq:angular momentum}), one finds
\be
	\frac{d\Omega}{dt} = \frac{n_*}{\tau_{{\rm syn},*}}
		\lp 1- \frac{\Omega}{n_*}\rp
		+ \frac{n_m}{\tau_{{\rm syn},m}}
		\lp 1- \frac{\Omega}{n_m}\rp,
		\label{eq:d omegap dt}
\ee
where the associated synchronization timescales are
\be
	\tau_{{\rm syn},*} \equiv \frac{\lambda}{3k_2\tau n_*}
			\lp \frac{M_p}{M_*} \rp
			\lp \frac{a_*}{R_p} \rp^3
			n_*^{-1},
\ee
and
\be
	\tau_{{\rm syn},m} \equiv \frac{\lambda}{3k_2\tau n_m}
			\lp \frac{M_p}{M_m} \rp^2
			\lp \frac{a_m}{R_p} \rp^3
			n_m^{-1},
\ee
and $\lambda=I_p/M_pR_p^2$ is the radius of gyration for the planet.

The orbital separations also change as described by the differential equations
\be
	\frac{d a_*}{dt} = -\frac{a_*}{\tau_{{\rm mig},*}}
	\lp 1- \frac{\Omega}{n_*}\rp,
	\label{eq:dastardt}
\ee
and
\be
	\frac{d a_m}{dt} = -\frac{a_m}{\tau_{{\rm mig},m}} 
	\lp 1- \frac{\Omega}{n_m}\rp
	\left[ 1-\frac{6}{5}\lp \frac{R_m}{a_m}\rp^2\right]^{-1},
	\label{eq:dasdt}
\ee
where the last term is due to synchronous spin of the moon and I assume $I_m=(2/5)M_mR_m^2$.
%This is a minor correction unless the moon is especially close to the planet.
For the change in the orbital orbital separations,
\be
	\tau_{{\rm mig},*}\equiv (6k_2\tau n_*)^{-1} \lp \frac{M_p}{M_*} \rp \lp \frac{a_*}{R_p} \rp^5 n_*^{-1},
\ee
and
\be
	\tau_{{\rm mig},m} \equiv (6k_2 \tau n_m)^{-1} \lp \frac{M_p}{M_m} \rp \lp \frac{a_m}{R_p} \rp^5 n_m^{-1}.
\ee
for the migration timescales. 

Besides the planet's spin and orbital separations, tides will also change the eccentricity of the orbits, which is governed by the equation \citep{Ogilvie14}
\be
	\frac{1}{e}\frac{de}{dt} = -\frac{3}{2}k_2\tau (18n_m-11\Omega) \frac{M_m}{M_p}\lp \frac{R_p}{a_m}\rp^5 n_m.
\ee
This happens on a roughly similar timescale to $\tau_{{\rm mig},m}$. Nevertheless, I assume circular orbits in this work because one of my main goals is to highlight the fact that tidal heating is possible even with no eccentricity.

Tidal heating can be present for circular orbits because when both the moon and star exert their tides on the planet, the planet is always forced to spin asynchronously with respect to each of them. To estimate the strength of this heating, first consider the total energy of the combined spinning planet plus gravitational interaction with a moon is
\be
	E_m = \frac{1}{2}I_p \Omega^2 - \frac{GM_pM_m}{2a_m}.
\ee
Taking the derivative of this expression results in
\be
	\dot{E}_m = I_p \Omega \frac{d\Omega}{dt} + \frac{GM_pM_m}{2a_m^2}\frac{da_m}{dt}.
\ee
Substituting the part of the torque due to the moon's tide on $d\Omega/dt$ and doing some algebra one finds
\be
	\dot{E}_m &=&  -\frac{\hat{\omega}_m}{2}N_m 
	= \frac{I_p\hat{\omega}_m^2}{4\tau_{{\rm syn},m}}
	\nonumber
	\\
	&=&\frac{3}{4}k_2\tau\hat{\omega}_m^2 M_m R_p^2  \lp\frac{M_m}{M_p} \rp \lp\frac{R_p}{a_m} \rp^3 n_m^2
\ee
This change of energy represents the maximum amount of heating possible on the planet due to asynchronous rotation with respect to the moon. Performing a similar set of arguments on the tidal forcing from the star results in
\be
	\dot{E}_* = -\frac{\hat{\omega}_*}{2}N_*
	= \frac{I_p\hat{\omega}_*^2}{4\tau_{{\rm syn},*}}
	= \frac{3}{4}k_2\tau \hat{\omega}_*^2M_*R_p^2 \lp\frac{R_p}{a_*} \rp^3 n_*^2.
	\nonumber
	\\
\ee
for the heating of the planet due to the star. {I take the total heating rate on the exoplanet to be the sum of the two tidal heating contributions, 
\be
	\dot{E} = \dot{E}_* + \dot{E}_m.
\ee
In detail, the tidal heating can occur at different locations depending on how the tides are damped, but this requires a more sophisticated treatment of the tides that is outside the scope of this work.}

\section{Characteristic Timescales and Analytic Evolution Solutions}
\label{sec:timescales}

The detailed evolution of the star-planet-moon system can be complicated because of the many parameters that can be varied. It is therefore helpful to consider some of the key timescales that will govern the evolution as well as solving a simpler set of evolution equations to provide some intuition on how the evolution will proceed.

\subsection{Orbital and Tidal Timescales}

First we consider the four key timescales for the synchronization and migration of the orbits that were identified in Section \ref{sec:equations}. Substituting physical values, these are
\be
	\tau_{{\rm syn},*} &=& 1.8\times10^{10}
	\lp \frac{k_2\tau/\lambda}{580\,{\rm s}}\rp^{-1}
	\lp \frac{M_p}{M_\oplus}\rp
	\lp \frac{M_*}{M_\odot} \rp^{-2}
	\nonumber
	\\
	&&\times\lp\frac{R_p}{R_\oplus} \rp^{-3}
	\lp \frac{a_*}{1\,{\rm AU}} \rp^{6}
	{\rm yr},
	\label{eq:taustar}
	\\
	\tau_{{\rm syn},m} &=& 3.7\times10^9
		\lp \frac{k_2\tau/\lambda}{580\,{\rm s}}\rp^{-1}
	\lp \frac{M_p}{M_\oplus}\rp
	\lp \frac{M_m}{M_{\rm M}} \rp^{-2}
	\nonumber
	\\
	&&\times\lp\frac{R_p}{R_\oplus} \rp^{-3}
	\lp \frac{a_m}{a_{\rm M}} \rp^6
	{\rm yr},
	\label{eq:tauc}
	\\
	\tau_{{\rm mig},*} &=& 2.5\times10^{17}
		\lp \frac{k_2\tau}{191\,{\rm s}}\rp^{-1}
	\lp \frac{M_p}{M_\oplus}\rp
	\lp \frac{M_*}{M_\odot} \rp^{-2}
	\nonumber
	\\
	&&\times\lp\frac{R_p}{R_\oplus} \rp^{-5}
	\lp \frac{a_*}{1\,{\rm AU}} \rp^{8}
	{\rm yr},
	\\
	\tau_{{\rm mig},m} &=& 2.5\times10^{11}
			\lp \frac{k_2\tau}{191\,{\rm s}}\rp^{-1}
	\lp \frac{M_m}{M_{\rm M}} \rp^{-1}
		\nonumber
	\\
	&&\times
		\lp\frac{R_p}{R_\oplus} \rp^{-5}
	\lp \frac{a_m}{a_\moon} \rp^{8}
	{\rm yr},
	\label{eq:as}
\ee
where $R_\oplus=6.37\times10^8\,{\rm cm}$ is the radius of the Earth, and I have used  $\lambda = 0.33$ and $k_2=0.3$, which are motivated by empirical measurements and fits to the shear modulus and stiffness of the present day Earth \citep{Williams94,Henning09,Ray12,Heller13,Driscoll15}. The time lag is set to $\tau=638\,{\rm s}$ \citep{Lambeck77,Neron97}, which for the current values of the Earth--Moon system gives a migration rate of the moon of $da_m/dt=3.8\,{\rm cm\,yr^{-1}}$ and spindown rate of the planet of $dP/dt = -2.2\,{\rm ms\,century^{-1}}$, where $P=2\pi/\Omega$ is the planet's spin period. As a check on the general framework used here, these rates are both in agreement with the currently measured values for the Earth and Moon.

At least for the specific values used in \mbox{Equations (\ref{eq:taustar})--(\ref{eq:as}),} the ordering of the timescales is $ \tau_{{\rm syn},m} \lesssim \tau_{{\rm syn},*} \lesssim \tau_{{\rm mig},m} \ll \tau_{{\rm mig},*}$. Thus for the Sun-Earth-Moon system, the synchronization of the Earth with the Moon's orbit is occurring the fastest of any of these processes. As emphasized in the discussion in Section \ref{sec:introduction}, where things potentially get interesting is when $\tau_{{\rm syn},*}\approx\tau_{{\rm syn},m}$, so that both the tides acting from the star and moon are impacting the planet on similar timescales. Setting $\tau_{{\rm syn},*}\approx\tau_{{\rm syn},m}$ using Equations (\ref{eq:taustar}) and (\ref{eq:tauc}), one can estimate that the typical separation between the planet and moon where this occurs is
\be
	a_m \approx 0.5
	\lp \frac{a_*}{0.3\,{\rm AU}} \rp
		\lp \frac{M_m}{M_{\rm M}} \rp^{1/3}
	\lp \frac{M_*}{0.5\,M_\odot} \rp^{-1/3}
	a_{\rm M}.
	\nonumber
	\\
\ee
Comparing this estimate for $a_m$ with Figure \ref{fig:timescales}, one can see that $\tau_{{\rm syn},*}\approx\tau_{{\rm syn},m}$ will naturally occur for low mass stars over a wide range of the parameter space where moons can remain bound to planets.

\subsection{Analytic Solutions without Orbital Migration}
\label{sec:analytic}

To get more intuition for what will happen to the system when $\tau_{{\rm syn},*}\approx\tau_{{\rm syn},m}$, I consider the simplified case when the orbital separations do not evolve. In other words, I make the approximation that $\tau_{{\rm mig},*}, \tau_{{\rm mig},m} \gg  \tau_{{\rm syn},*}, \tau_{{\rm syn},m}$. It is found for this case that a particularly simple analytic solution is possible for the planet's spin evolution.

Assuming that $a_m$ and $a_*$ are constant, Equation~(\ref{eq:d omegap dt}) becomes a differential equation simply in $\Omega(t)$. Integrating this then results in
\be
	\Omega(t) = \Omega_{\rm eq} + \lp \Omega_0 - \Omega_{\rm eq} \rp \exp \lp - t/\tau_{\rm syn} \rp,
	\label{eq:chi}
\ee
where $\Omega_0\equiv \Omega(t=0)$ is the initial spin frequency, $\Omega_{\rm eq}$ is the equilibrium spin frequency as $t\rightarrow\infty$, and the total synchronization time of the planet is defined to be
\be
	\tau_{\rm syn} \equiv \frac{\tau_{{\rm syn},m}\tau_{{\rm syn},*}}{\tau_{{\rm syn},m}+\tau_{{\rm syn},*}}.
\ee
The easiest way to estimate $\Omega_{\rm eq}$ is to just set $d\Omega/dt=0$ with Equation~(\ref{eq:d omegap dt}), and then solve for $\Omega$, resulting in.
\be
	\Omega_{\rm eq} \equiv \frac{\tau_{{\rm syn},m}n_*+\tau_{{\rm syn},*}n_m}{\tau_{{\rm syn},m}+\tau_{{\rm syn},*}}.
	\label{eq:omega_eq}
\ee
Equation~(\ref{eq:chi}) shows that the planet's spin just exponentially decays on a timescale $\tau_{\rm syn}$ to the final equilibrium spin of  $\Omega_{\rm eq}$.

In general, this solution is not exactly correct because $a_m$ can also potentially evolve on similar timescales as shown by Equation (\ref{eq:as}). This in turn changes $\tau_{{\rm syn},*}$ and $\tau_{{\rm syn},m}$, so that they are not constant as assumed to derive \mbox{Equation (\ref{eq:chi}).} An example where this fails is for the Sun-Earth-Moon system, where the migration of the Moon cannot be ignored.

 \begin{figure}
\begin{center}
  \includegraphics[width=0.47\textwidth,trim=0.0cm 0.0cm 0.0cm 0.0cm]{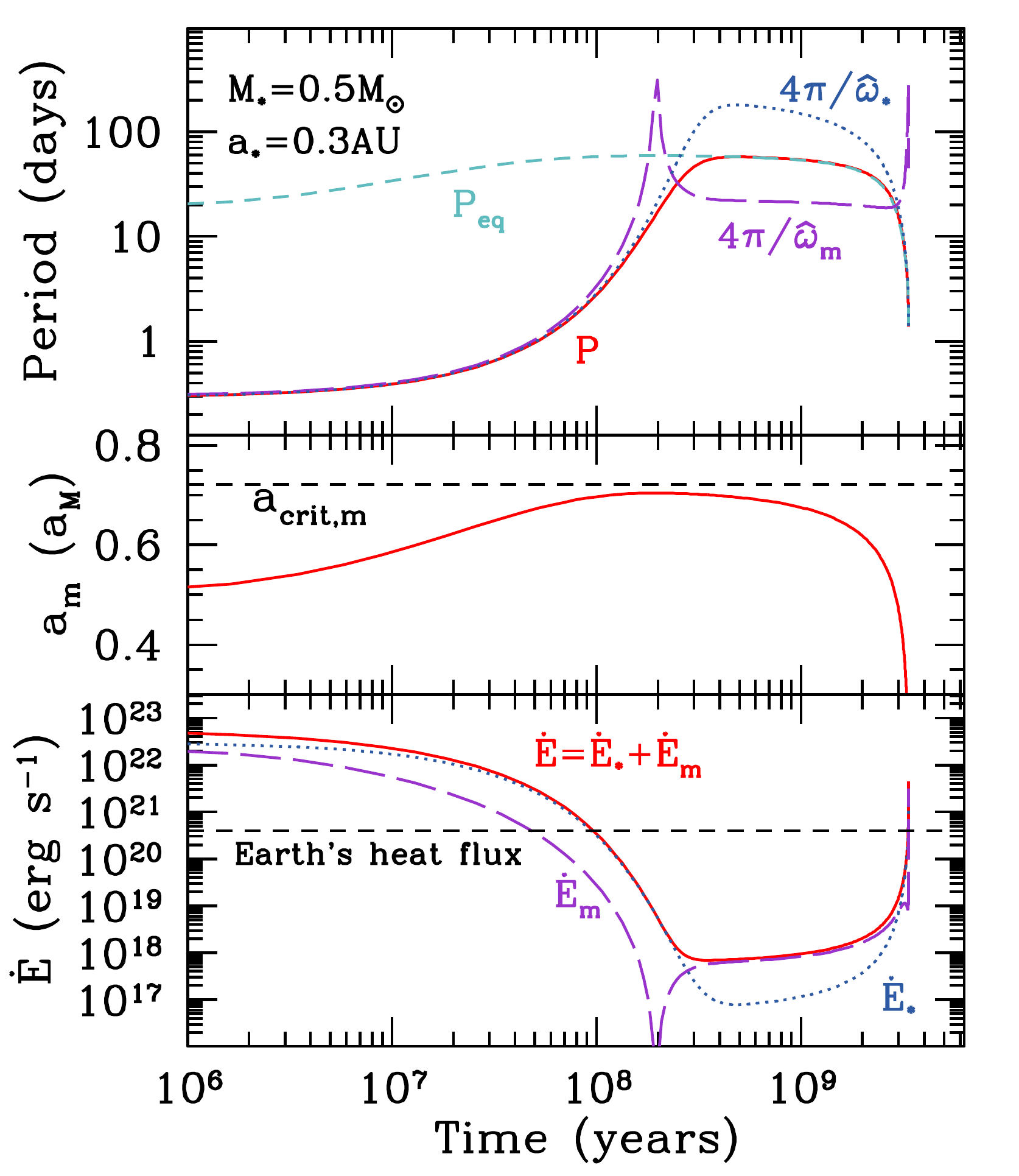}
  \end{center}
\caption{An example time evolution of the star-planet-moon system, using the parameters $M_*=0.5\,M_\odot$, $M_p=M_\oplus$, and $M_m=M_\moon$, with initial orbital separations $a_*=0.3\,{\rm AU}$ and $a_m=0.5\,a_\moon$, and an initial spin period for the planet of $P_0=7\,{\rm hrs}$. The top panel summarizes the main periods of the system $P_{\rm eq}=2\pi/\Omega_{\rm eq}$ (dashed turquoise line), $P=2\pi/\Omega$ (solid red line), $4\pi/\hat{\omega}_*$ (dotted blue line), and $4\pi/\hat{\omega}_m$ (long dashed purple line). The middle panel is the moon's orbital separation (red solid line) in comparison to the critical radius $a_{{\rm crit},m}$ (dashed black line) at which the moon would be tidally stripped by the star from Equation (\ref{eq:roche}). The bottom panel compares tidal heating rates, which includes the heating from the star $E_*$ (dotted blue line), the moon $E_m$ (long dashed purple line), and the total heating (solid red line). Also plotted is the current heat flux coming up through the Earth of $\approx 4\times10^{20}\,{\rm erg\,s^{-1}}$ (dashed black line).}
\epsscale{1.0}
\label{fig:fiducial}
 \end{figure}

Nevertheless, the full evolutions of the star-planet-moon system will demonstrate that the concept of $\Omega_{\rm eq}$ plays an important role. In particular, note that from Equation (\ref{eq:omega_eq}) it is apparent that $\Omega_{\rm eq}$ cannot exactly equal either $n_*$ or $n_m$. This means that there will be tidal forcing on the planet by both the star and the moon, and tidal heating will always play some role in a star-planet-moon system.

\section{Full Time Evolved Solutions}
\label{sec:solutions}

Now that the main background has been covered, I consider the full evolution of the star-planet-moon system. This is solved by numerically integrating forward in time Equations (\ref{eq:d omegap dt}), (\ref{eq:dastardt}), and (\ref{eq:dasdt}), which are three coupled differential equations in the dependent variables $\Omega$, $a_*$, and $a_m$, respectively.

\subsection{Example Evolution}

It is helpful to first just focus on one fiducial example that exemplifies the main features on the solutions, which is presented in Figure \ref{fig:fiducial}. For this specific case, I use $M_*=0.5\,M_\odot$, $M_p=M_\oplus$, and $M_m=M_\moon$ (in all further examples, $M_p$ and $M_m$ use these values for simplicity). The initial orbital separations are $a_*=0.3\,{\rm AU}$ and $a_m=0.5\,a_\moon$. Finally, the initial spin of the planet must be chosen, in this case I use $P_0=2\pi/\Omega_0=7\,{\rm hrs}$ (note that the Earth is generally thought to have had an initial spin period of roughly $6\,{\rm hrs}$).

From this example, a number of important features are seen that will inform our more detailed parameter survey below. First, in the upper panel of Figure \ref{fig:fiducial}, I summarize the key periods of the system. These are the actual spin period of the planet $P=2\pi/\Omega$ (solid red line), the equilibrium spin period related to the analytic solutions from Section \ref{sec:analytic} $P_{\rm eq}=2\pi/\Omega_{\rm eq}$ (dashed turquoise line), $4\pi/\hat{\omega}_*$ (dotted blue line), and $4\pi/\hat{\omega}_m$ (long dashed purple line). These latter two periods are the periods of the tidal forcing from the star and moon, respectively, with an extra factor of 2 in each case to cancel the factor of 2 from the two tidal bulges. Also, for these periods the absolute value is plotted since they can be either negative or positive. At early times we see that $P \approx 4\pi/\hat{\omega}_* \approx 4\pi/\hat{\omega}_m$. This is because the fast initial spin of the planet dominates setting the tidal frequencies. Furthermore, both of the toques are negative, and the planet is spinning down. At a time of $\approx2\times10^8\,{\rm yrs}$, the planet has spun down sufficiently that $\Omega<n_m$, which causes the torque from the moon to instead want to spin the planet up. This switch in the sign of the moon's torque can be seen in the plot of $4\pi/\hat{\omega}_m$, which shows a cusp at this time. Now the combined torques of the moon (spinning the planet up) and the star (spinning the planet down) push the planet's spin toward $P\approx P_{\rm eq}$.

\begin{figure}
\begin{center}
  \includegraphics[width=0.5\textwidth,trim=0.0cm 4.0cm 0.0cm 5.0cm]{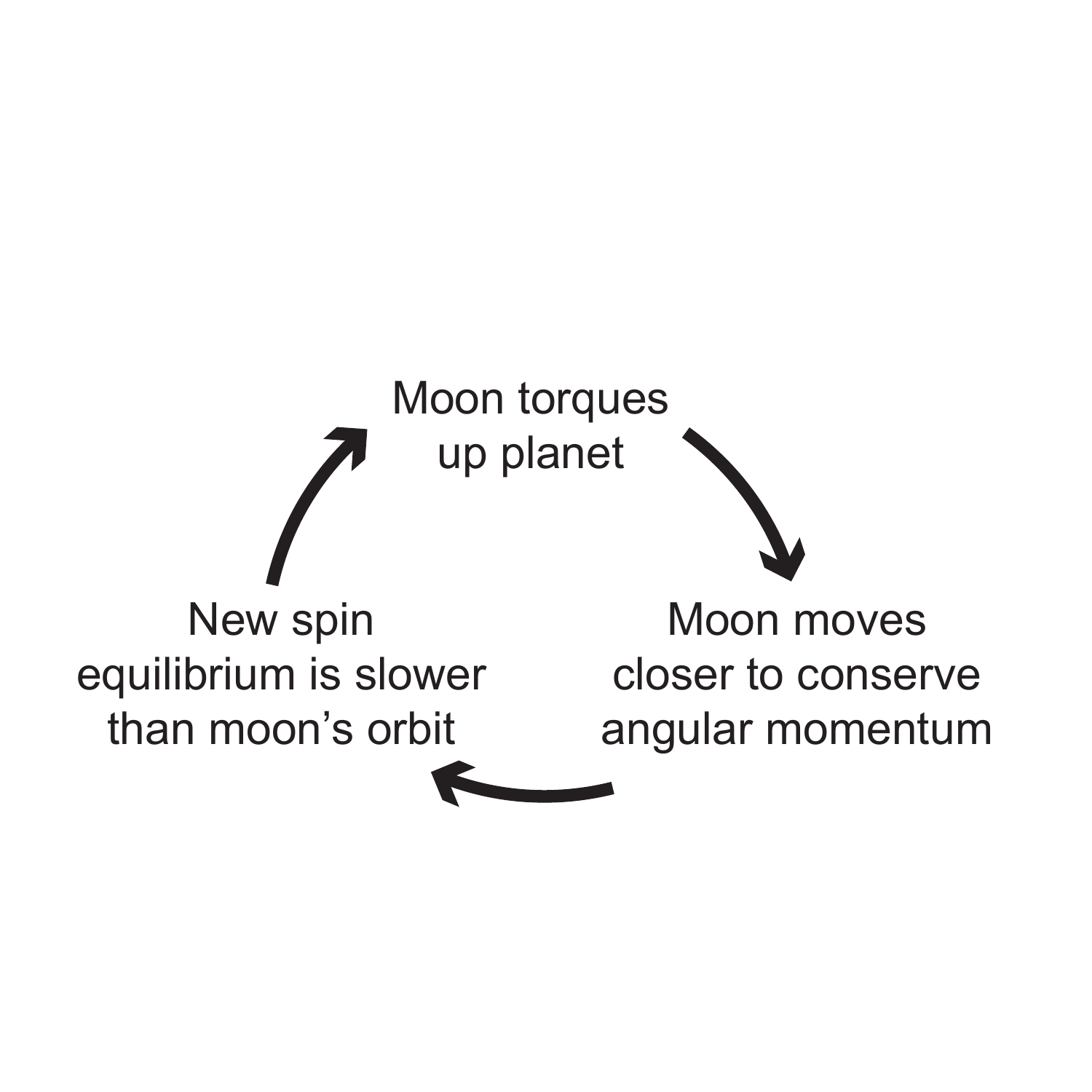}
  \end{center}
  \caption{Although the planet can spin down until $\Omega\approx \Omega_{\rm eq}$, this is not a stable equilibrium. Because $n_m>\Omega_{\rm eq}>n_*$, this means that (1) the moon torques up the planet, which then (2) moves the moon closer to conserve angular momentum, so that (3) at the new spin equilibrium again $n_m>\Omega_{\rm eq}$ because $\Omega_{\rm eq}>n_*$. Thus this loops continue and the moon moves in toward the planet until it is disrupted.}
  \label{fig:unstable_diagram}
\end{figure}

Even though the planet has reached this spin equilibrium, the story is not over because this equilibrium is not stable. As shown by Equation (\ref{eq:omega_eq}), if spin equilibrium is reached then the planet is not tidally locked with either the moon or the star, and thus the tidal forces still persist. Given the relative frequencies, in this situation the ordering is $n_m>\Omega_{\rm eq}>n_*$, and therefore the planet will be spun down by the star but spun up by the moon. Thus we find that an equilibrium is reached, but the equilibrium is not stable as summarized in Figure \ref{fig:unstable_diagram}. Since $n_m>\Omega_{\rm eq}$, the moon torques up the planet. This then moves the moon closer to the planet to conserve angular momentum. With the new orbital separation, once again $n_m>\Omega_{\rm eq}$, and thus the process repeats until the moon moves close enough to the planet to be disrupted.

This is seen in the middle panel of Figure \ref{fig:fiducial}, which plots the evolution of the moon's orbital separation (red solid line). At early times, $a_m$ increases as the planet spins down and donates its angular momentum to the moon's orbit until leveling off when $P\approx P_{\rm eq}$. This does not last though because of the unstable situation and eventually $a_m$ comes crashing back to the planet. Also plotted in Figure \ref{fig:fiducial} is the critical radius for the moon to be lost to the parent star $a_{{\rm crit},m}$ (black dashed line) given by Equation (\ref{eq:roche}). In this particular case, $a_m<a_{{\rm crit},m}$ for the entire evolution, but if this critical radius is ever exceeded, one should expect the moon to be tidally stripped from the planet.

Finally, in the bottom panel of Figure \ref{fig:fiducial}, I summarize the tidal heating rates.  At early times, this is dominated by the star, and at late times the moon. But in either case, it must remain non-zero because the combined tides always make sure the planet is asynchronous with the orbits of both the star and moon. Also plotted is the heat flux of $\approx 4\times10^{20}\,{\rm erg\,s^{-1}}$ that is currently emanating through the Earth \citep[black dashed line,][]{Dye12} due to a combination of radioactivity, latent heat, and heat left over from the Earth's formation. This demonstrates that the tidal heating is actually similar or exceeds this value for a timescale of $\approx10^8\,{\rm yrs}$, and thus one might expect such a planet to have tectonic activity similar to the Earth (or perhaps even Io, which has an even greater heating rate) during this time. Although compared to the timescale necessary for life to develop, this appears too short (albeit, such a timescale is obviously very uncertain).

\subsection{Is the Moon Disrupted or Lost?}

 \begin{figure}
\begin{center}
  \includegraphics[width=0.47\textwidth,trim=0.0cm 0.0cm 0.0cm 0.0cm]{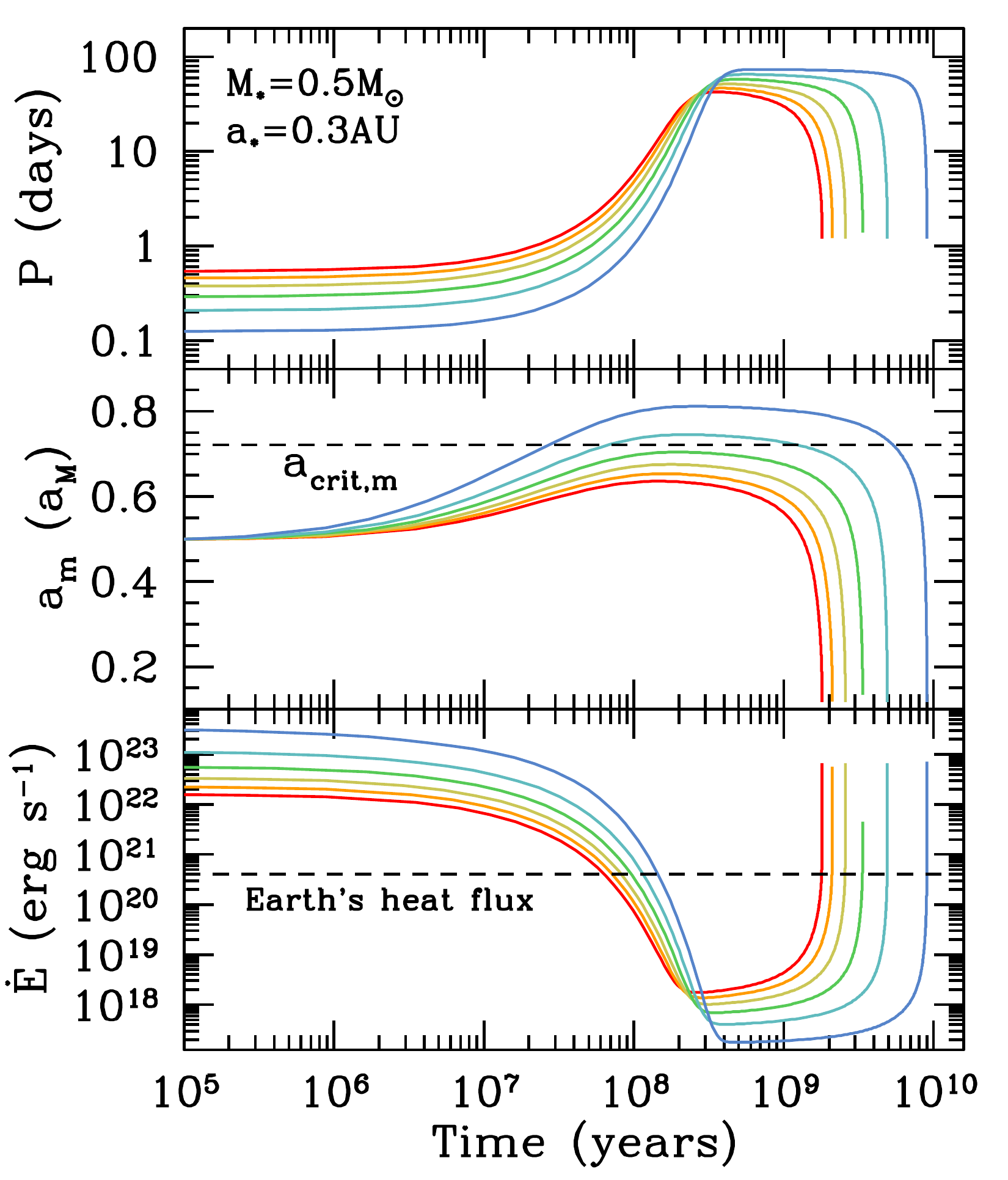}
  \end{center}
\caption{Similar to Figure \ref{fig:fiducial}, but varying the initial spin period $P_0$ of the planet with values of $3\,{\rm hrs}$ (blue curves), $5\,{\rm hrs}$ (turquoise curves), $7\,{\rm hrs}$ (green curves), $9\,{\rm hrs}$ (yellow curves), $11\,{\rm hr}$ (orange curves), and  $13\,{\rm hrs}$ (red curves). All other parameters are the same as in Figure \ref{fig:fiducial}. This demonstrates when $P_0=7\,{\rm hrs}$, the moon will likely be disrupted, but if the initial spin is shortened to $P_0=5\,{\rm hrs}$, then the moon is likely stripped by the star.}
\epsscale{1.0}
\label{fig:planet spin}
 \end{figure}

In the example from the previous section and in Figure \ref{fig:fiducial}, it appears the moon will be forced to migrate into the planet rather than expelled by the star. But this example also showed that the moon was fairly close to being removed by the star if only the moon migrated out a little further (compare the red solid line and black dashed lines in the middle panel of \mbox{Figure \ref{fig:fiducial}).}

To better explore what controls the fate of the moon, in Figure \ref{fig:planet spin} we plot the evolution of the star-planet-moon system for a variety of different initial spins for the planet. The initial spins vary from from periods of $3\,{\rm hrs}$ (blue curves) to $13\,{\rm hrs}$ (red curves). This demonstrates that the initial spin of the planet can have a dramatic affect on the fate of the moon. The reason is that the moon's migration is driven by the extraction of angular momentum from the planet's spin, and the more spin the planet has, then the further out the moon will migrate. This dependency on $P_0$ is not emphasized in the work of \citet{Sasaki12}, which used a constant value of $P_0=6\,{\rm hrs}$ for Earth-like planets {(although see some of the discussion in \citealp{Sasaki14}).}

Whether the moon is disrupted or lost can be addressed with some simple arguments. The initial angular momentum in the planet's spin is $2\pi \lambda M_p R_p^2/P_0$. If all of this angular momentum goes into the moon's orbit, with angular momentum $M_m(GM_pa_m)^{1/2}$, then the orbital separation of the moon would be
\be
	a_m = \frac{4\pi^2\lambda^2M_p R_p^4}{GM_m^2P_0^2}.
\ee
Equating this to $a_{{\rm crit},m}$ given by Equation (\ref{eq:roche}) and solving for $P_0$ provides the critical initial period for the planet
\be
	P_{{\rm crit},0} &=& \frac{2\pi \lambda M_p^{1/2}R_p^2}{(Gfa_*)^{1/2}M_m} \lp\frac{3M_*}{M_p} \rp^{1/6},
	\nonumber
	\\
	&=& 5.7 \lp \frac{M_p}{M_\oplus}\rp^{1/3}
	\lp \frac{M_m}{M_{\rm M}} \rp^{-1}
	\lp \frac{R_p}{R_\oplus} \rp^2
	\nonumber
	\\
	&&\times \lp\frac{a_*}{0.3\,{\rm AU} }\rp^{-1/2}
	\lp\frac{M_*}{0..5\,M_\odot} \rp^{1/6}
	{\rm hrs}.
	\label{eq:p crit}
\ee
This matches fairly closely the critical value for the initial period found from the evolutions in Figure \ref{fig:planet spin}.

\begin{figure}
\begin{center}
  \includegraphics[width=0.47\textwidth,trim=0.0cm 0.0cm 0.0cm 0.0cm]{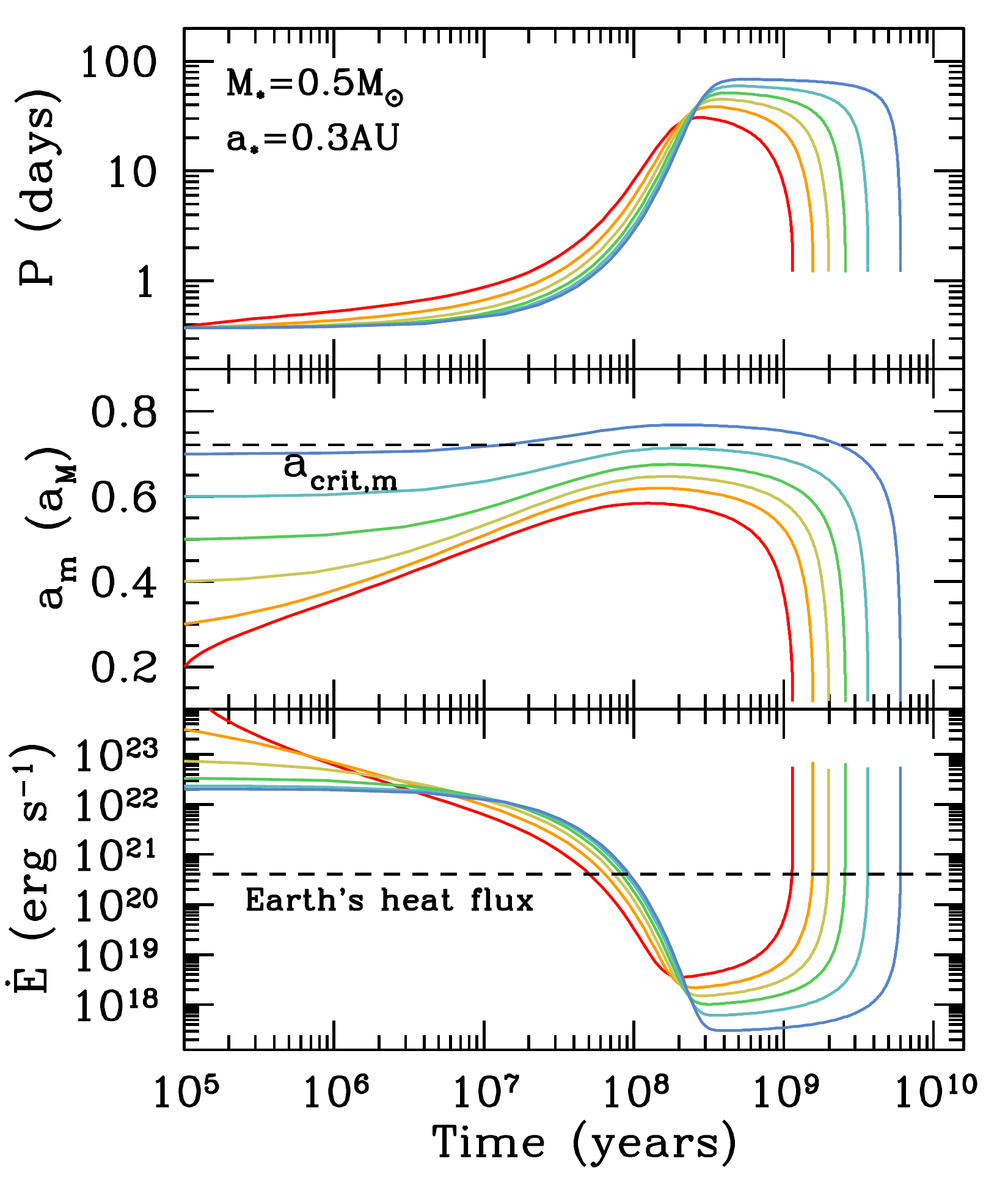}
  \end{center}
\caption{Similar to Figure \ref{fig:fiducial}, but with $P_0=9\,{\rm hrs}$ and  varying the orbital separation of the moon over the values $0.7a_\moon$ (blue curves), $0.6a_\moon$ (turquoise curves), $0.5a_\moon$ (green curves), $0.4a_\moon$ (yellow curves), $0.3a_\moon$ (orange curves), and  $0.2a_\moon$ (red curves). All other parameters are the same as in Figure \ref{fig:fiducial}. This demonstrates that for  $0.7a_\moon$, the moon is sufficiently far away to begin with that only with a small amount of migration it will be stripped by the star.}
\epsscale{1.0}
\label{fig:am}
 \end{figure}

In detail, this critical period does not exactly hold. This is because it matters how much angular momentum the moon initially has as well. For example, in Figure \ref{fig:am} for each of the evolutions the initial spin period of the planet is set to $P_0=9\,{\rm hrs}$ (in Figure \ref{fig:planet spin}, this $P_0$ was found to be sufficiently long that the moon is expected to migrate back into the planet). But varying the moon's initial orbital separation from $0.2a_\moon$ to $0.7a_\moon$ demonstrates that if the moon is initially sufficiently far, then it will be easier to strip. This effect is not captured by Equation (\ref{eq:p crit}).

{Another physical mechanism that will play a role in determining the fate of the moon is the role of tidal resonances. Of particular interest may be the evection resonance, where the moon's perihelion precession rate becomes equal to the orbital period of the planet (e.g., \citealp{Touma98}). This can alter the moon's orbit dramatically, but it depends in detail on the rate of the orbital evolution and the width of the resonance. Due to these complications, I save a more detailed study of these processes for future work.}

\subsection{Fate of the Disrupted Moon}

In the cases where the moon is forced to tidally migrate back toward the planet, there are likely two potential outcomes: (1) the moon is tidally disrupted, or (2) the moon directly impacts the planet. Here it is argued that at least for a system similar to the Earth-Moon, tidal disruption is more likely.

The moon will migrate inward until its radius hits the Roche lobe, i.e., the equipotential surface where material will no longer be gravitationally bound to the moon. As long as $M_m/M_p\ll1$, this can be approximated by the condition \citep{Frank02}
\be
	R_m = 0.462 \lp 1+\frac{M_p}{M_m} \rp^{-1/3} a_t \approx 0.462 \lp \frac{M_m}{M_p} \rp^{1/3} a_t,
	\nonumber
	\\
	\label{eq:disruption}
\ee
where $a_t$ denotes the semi-major axis when tidal disruption occurs. At the moment of disruption, the L1 Lagrange point is located a distance $d_t \approx 0.7(M_m/M_p)^{1/3}a_t$ from the center of mass of the moon. Thus, for tidal disruption to occur, then $a_t\gtrsim R_p+d_t$, otherwise the planet and moon are too close together and direct impact will occur instead. Rewriting this inequality by making use of Equation (\ref{eq:disruption}),
\be
	2.16\lp \frac{M_p}{M_m} \rp^{1/3} \gtrsim \frac{R_p}{R_m} + 1.51.
\ee
Simply using the properties of the Earth and Moon, the left side is $9.35$ while the right side is $5.19$, thus the result is tidal disruption. Another way to understand this is to multiple both sides of this expression by $R_m/R_p$, which results in
\be
	\lp \frac{\langle\rho_p\rangle}{\langle\rho_m\rangle} \rp^{1/3} \gtrsim 0.46 + 0.70\frac{R_m}{R_p},
	\label{eq:density}
\ee
where $\langle\rho_p\rangle$ and $\langle\rho_m\rangle$ are the average density of the planet and moon, respectively. Since the righthand side of \mbox{Equation (\ref{eq:density})} has a value between $0.46$ to $1.16$ (since $R_m<R_p$), this demonstrates that as long as the density of the planet is similar or greater than the moon, then tidal disruption is expected. Conversely, direct impact only occurs when the density of the planet is much less than the moon \citep[similar arguments were presented in][in a somewhat different context]{Metzger12}.

\begin{figure}
\begin{center}
  \includegraphics[width=0.47\textwidth,trim=0.0cm 0.0cm 0.0cm 0.0cm]{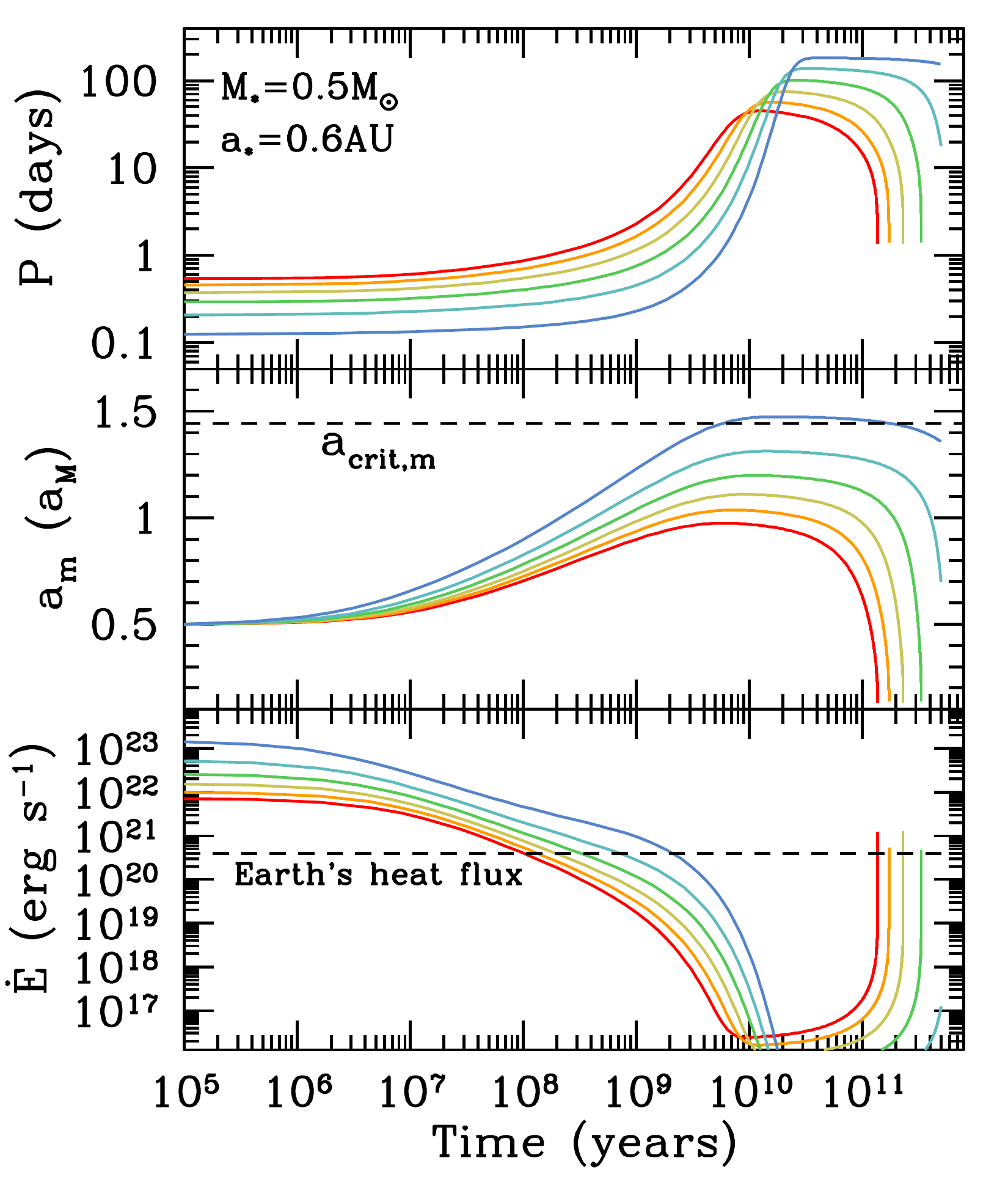}
  \end{center}
\caption{Same as Figure \ref{fig:planet spin}, but with $a_*=0.6\,{\rm AU}$. This greatly extends the evolution to longer timescales and the moon gets tidally disrupted for a larger range of initial spin periods for the planet $P_0$.}
\epsscale{1.0}
\label{fig:large astar}
 \end{figure}
 
      \begin{figure}
\begin{center}
  \includegraphics[width=0.47\textwidth,trim=0.0cm 0.0cm 0.0cm 0.0cm]{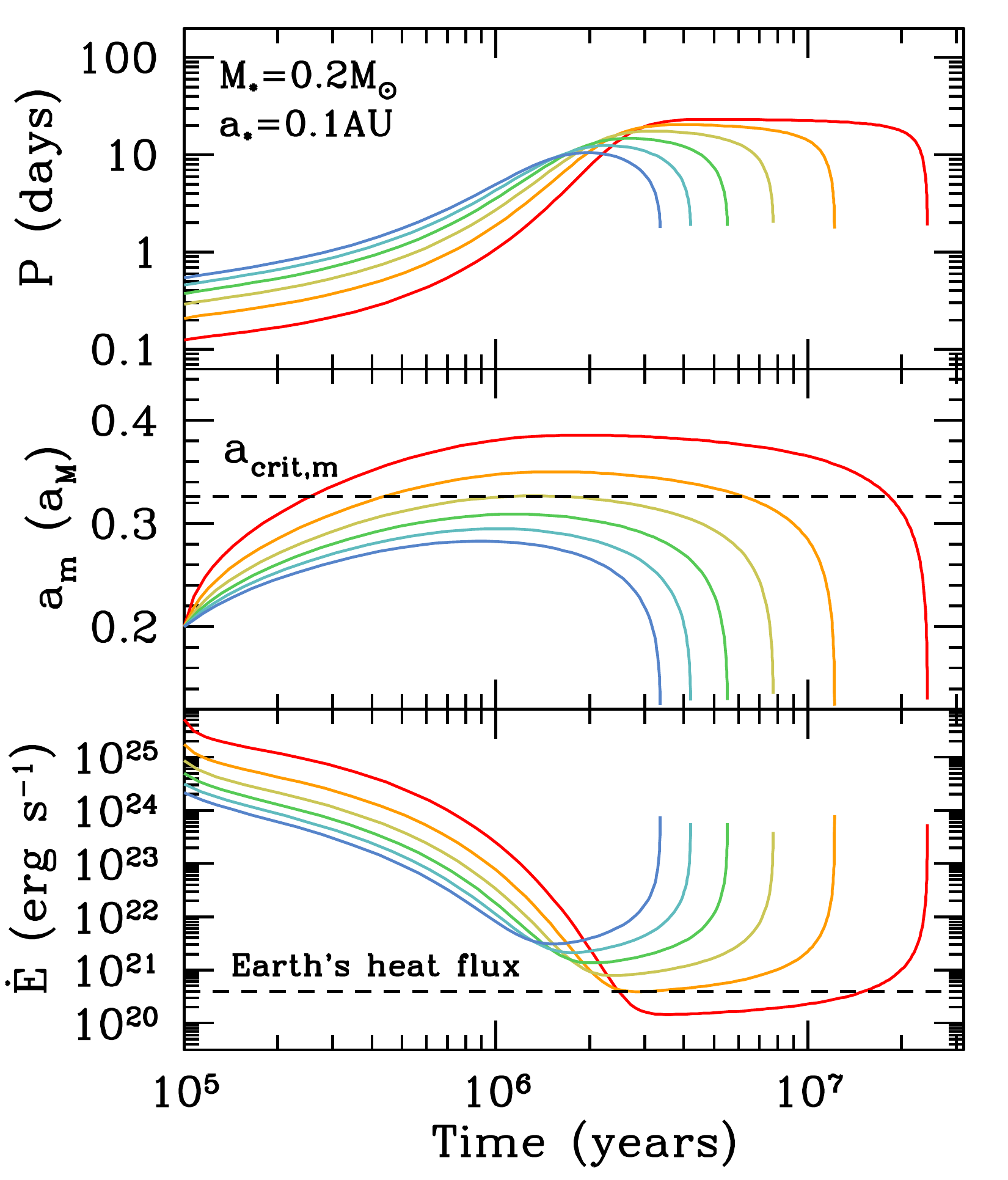}
  \end{center}
\caption{Same as Figure \ref{fig:planet spin}, but with $M_*=0.2\,M_\odot$ and $a_*=0.1\,{\rm AU}$. The small separation greatly speeds up the evolution.}
\epsscale{1.0}
\label{fig:small star}
 \end{figure}

Following disruption, the moon will likely form some sort of ring-like structure around the planet. This may indicate that some rocky planets around low mass stars should be expected to have circumplanetary rings. To get some idea of what such rings may look like, consider that at the moment of disruption at a separation $a_t$, the orbital angular momentum of the moon is
\be
	J_t &=& [G(M_p+M_m)a_t]^{1/2}M_m
	\nonumber
	\\
	&\approx& 1.47 [G(M_p+M_m)R_m]^{1/2}M_p^{1/6}M_m^{5/6}.
\ee
where for the second expression I have used Equation (\ref{eq:disruption}). For the values of the Earth-Moon system, this gives an angular momentum of $J_t\approx6\times10^{40}\,{\rm erg\,s}$. This same amount of angular momentum should be roughly stored in the resulting rings, which would be given by
\be
	J_r \approx (GM_pR_r)^{1/2}M_r,
\ee
where $R_r$ and $M_r$ are the radius and mass of the rings, respectively. The actual mass that goes into the rings would depend on the details of the ring dynamics, but even if $\approx10\%$ of the moon's mass went into the rings it would imply a radius of $R_r\approx2\times10^{11}\,{\rm cm}$ (and an even larger radius if the ring mass is smaller). {Such a large radius cannot be maintained for the rings. Material interior to the so-called fluid Roche limit will remain in rings, where the fluid Roche limit is given by \citep{Murray99}
\be
	R_{\rm FRL} = 2.46 R_p\lp \frac{\langle\rho_p\rangle}{\langle\rho\rangle} \rp^{1/3},
\ee
where $\langle\rho\rangle$ is the average density of the material that makes up the rings. This results in typical ring radii of $\approx2\times10^9\,{\rm cm}$. Searching for such rings \citep[using the methods described in, for example,][]{Barnes04,Ohta09,Zuluaga15} may verify that processes as discussed here took place in a specific system. Exterior to the fluid Roche limit, material can coalesce into a new moon that is some fraction of the mass of the moon before. This new moon could then migrate inward again and the process repeat. This suggests that the planet may go through phases where it alternatively has a moon or rings, which has actually been suggested to be the case for Mars and Phobos \citep{Hesselbrock17}. More work is needed to better understand the evolution and duty cycle of such rings.}

\subsection{Exploration of More Evolutions}
 
The above discussions spell out some of the general features of the evolution, but there are many parameters that can be varied for a star-planet-moon system. Thus here I highlight some other example evolutions to provide some sense to the diversity of potential results. In each example, I consider a range of values for $P_0$, since this factor has proven to be key in determining the moon's fate.

In the above examples, I set $a_*=0.3\,{\rm AU}$ for the initial separation so that the planet would be within the habitable zone, but what if the planet is much further out? In Figure \ref{fig:large astar}, I consider such a case with $a_*=0.6\,{\rm AU}$. The two main results of this change are that (1) the moon avoids being tidally stripped by the star for a larger range of initial spin periods for the planet and (2) it takes a much longer timescale for the moon to migrate back into the planet. For the first case, this is simply because the ability of the star to strip the moon depends most strongly on the separation, as shown by Equation (\ref{eq:roche}). For the second case, this is because the tidal timescales depend on a high power of the separation. In addition, tidal heating also stays strong for a longer period of time, exceeding the current heat flux on Earth for up to $\sim10^9\,{\rm yrs}$.

\begin{figure}
\begin{center}
  \includegraphics[width=0.47\textwidth,trim=0.0cm 0.00cm 0.0cm 0.0cm]{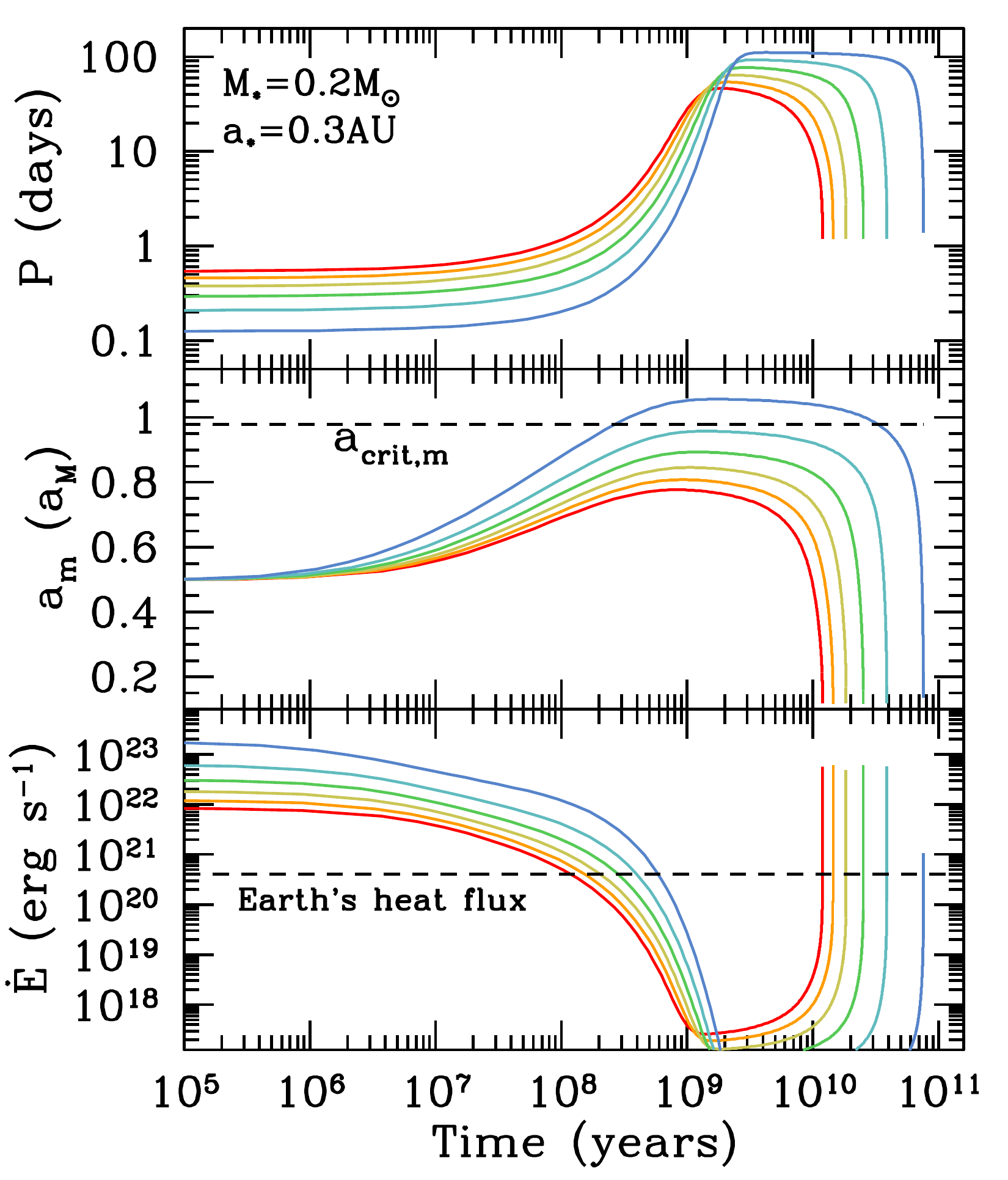}
  \end{center}
\caption{Same as Figure \ref{fig:planet spin}, but with $M_*=0.2\,M_\odot$ and $a_*=0.3\,{\rm AU}$. In comparison to Figure \ref{fig:small star}, the larger separation dramatically lengthens the evolution.}
\epsscale{1.0}
\label{fig:small star large astar}
 \end{figure}

 In Figure \ref{fig:small star}, I consider the case of a smaller star with $M_*=0.2\,M_\odot$, which requires an initial separation of $a_*=0.1\,{\rm AU}$ for the planet to be near the habitable zone. Although the mass of the star is smaller, the planet's orbital separation plays a much stronger role in setting the tidal interaction timescales. Thus, the evolution occurs much faster in this case with the fate of the moon being decided within $\sim10^6\,{\rm yrs}$. In comparison, in Figure \ref{fig:small star large astar}, just by tripling the initial separation to $a_*=0.3\,{\rm AU}$, now the evolution can occur for $10^{10}\,{\rm yrs}$ or longer.

\section{Discussion and Conclusions}
\label{sec:conclusions}

In this work, I have investigated the tidal interactions of star-planet-moon systems. In comparison to the broader work of \citet{Sasaki12}, this study has a more specific focus on Earth-Moon-like systems around low mass stars, which is motivated by recent surveys finding planets in the habitable zones of these stars. Furthermore, since these habitable zones are relatively close to the star, the tidal interactions between the planet and star are naturally comparable to the interactions between the planet and moon. I especially highlight the role of the initial spin period of the planet $P_0$ in determining the fate of the moon, and I use a constant $\tau$ formalism rather than a quality factor for assessing the impact of the tides. 

Solving for the time evolution of these systems, my main conclusions are as follows.
\begin{itemize}
\item The combined tidal interactions cause the moon to eventually be stripped by the star or migrate back toward the planet.
\item Which of these fates befall the moon depends sensitively on the initial spin period of the planet $P_0$, with a small spin making it more likely for the moon to be stripped.
\item In cases where the moon migrates into the planet, the moon will be tidally disrupted rather than directly impact the planet because of the relatively similar densities of the rocky planet and moon. {This may produce rings around rocky planets.}
\item The combined tidal interactions force the planet to always spin asynchronously with respect to the both the moon and star, often generating an amount of tidal heating similar to the current heat flux coming up through the Earth for up to $\sim10^9\,{\rm yrs}$.
\item The overall evolution of this system until the time the moon is stripped or disrupted depends on the initial separation of the planet and star, and can be very greatly from less than $10^6\,{\rm yrs}$ to greater than $10^{10}\,{\rm yrs}$.
\end{itemize}
In the future, as extrasolar moons are inevitable discovered, the formalism presented here can be used to better understand the lifetime and fate of these moons (although in some cases additional planetary bodies may also impact the moon, something outside the scope of this work). If instead the presence of rings around rocky planets orbiting low mass stars is found, it would provide evidence that processes as described here have occurred in specific systems. Alternatively, if no rings or moons are ever present, it could indicate that the moons are stripped because of the short initial spin when the planet is formed, providing insight into the planet formation process. Further calculations are needed to understand the details of these rings, how they evolve, and how long they should be present.

It is interesting to ponder the implications of the long timescale found for the loss of the moon through tidal disruption or stripping. One might assume that the fate of a moon would be determined relatively early during the formation process of the solar system and planet. But in fact, without fine tuning the parameters much, a moon could orbit a planet for well over $10^{10}\,{\rm yrs}$ before being tidally disrupted or lost from the planet completely! If such a planet could harbor life, this would presumably be sufficiently long for an advanced civilization to develop, only to be subject to a catastrophic event. Nevertheless, as we have seen over the last century, technology advances quickly. And although our species still seems to be sorting out issues more local to home, one might hope that in a relatively short amount of time in comparison to astrophysical timescales, an advanced society may be able to overcome such a unique challenge.

\acknowledgments
I thank Johanna Teske for suggestions on a previous draft of this manuscript, and Jason Barnes for answering my questions about his work. {I also thank Konstantin Batygin, Alexandre Correia, Michael Efroimsky, Jim Fuller, and Valeri Makarov for their feedback.}

\end{document}